\documentclass[aps,prd,twocolumn,superscriptaddress,nofootinbib,showpacs,preprintnumbers]{revtex4}
\usepackage[english]{babel}

\begin{document}
\preprint{IFF-RCA-09-04}
\title{Thermal radiation from Lorentzian traversable wormholes.}

\author{Prado Mart\'{\i}n-Moruno}
\email{pra@imaff.cfmac.csic.es} \affiliation{Colina de los
Chopos, Instituto de F\'{\i}sica Fundamental, \\
Consejo Superior de Investigaciones Cient\'{\i}ficas, Serrano 121,
28006 Madrid, Spain}
\author{Pedro F. Gonz\'{a}lez-D\'{\i}az}
\email{p.gonzalezdiaz@imaff.cfmac.csic.es } \affiliation{Colina de
los
Chopos, Instituto de F\'{\i}sica Fundamental, \\
Consejo Superior de Investigaciones Cient\'{\i}ficas, Serrano 121,
28006 Madrid, Spain}

\begin{abstract}
In this paper we show that, analogously to as it occurs for black
holes, there exist three well-defined laws for Lorentzian wormhole
thermodynamics and that these laws are related with a thermal
phantom-like radiation process coming from the wormhole throat. It
is argued that the existence of wormholes could be manifested by
means such a radiation. These results are obtained by analyzing
the Hayward formalism of spherically symmetric solutions
containing trapping horizons, the phenomenon of phantom accretion
onto wormholes and the development of phantom thermodynamics.
\end{abstract}

\pacs{04.62.+v, 04.70.Dy}

\keywords{Lorentzian dynamic wormholes, thermodynamics, thermal
radiation.}

\maketitle

\section{Introduction.}
In a broad sense the term wormhole was coined by Misner and Wheeler \cite{Misner:1957mt} (see also Wheeler \cite{Wheeler:1955zz}) even though some other authors used the concept quite before, including Flamm \cite{Flamm}, Weyl \cite{Weyl} and the so-called Einstein-Rosen bridge \cite{Einstein:1935tc}. The research on wormholes continued later with the contributions of authors like e. g. Zipoy \cite{Zipoy}, Ellis \cite{Ellis:1973yv}, Bronnikov \cite{Bronikov} and Clement \cite{Clement:1983fe}. Wormholes, which were both traversable and stable, were
considered and studied by Morris and Thorne \cite{Morris:1988cz}, who assumed that there can be traversable wormhole solutions and worked out some inferences about their properties. Throughout this paper we will use the term wormhole to mean a stable and traversable wormhole, unless otherwise stated. The fascinating
property of these wormholes lies on the fact that besides describing
short-cuts between two universes or two separate regions of one
universe, if one lets one of their mouths to move
relative to the other, they can be converted into time machines \cite{Morris:1988tu}. The main problem with such wormholes is, however, that one of
the requirements that they must fulfill in order to be traversable
from one mouth to the other is the absence of an event horizon,
and this in turn implies that wormholes must be supported by some
kind of the so-called ``exotic matter'', characterized by
violating the null energy condition. Even though this pathology
has not caused the total neglect of these solutions, it has
nevertheless led to the study of special cases where the necessary
amount of this exotic matter could be minimized (see chapter 15 of \cite{Visser}).

Recent astronomical data \cite{Mortlock:2000zu} indicate that the Universe could be
dominated by a stuff which violates the null energy condition,
dubbed phantom energy \cite{Caldwell:1999ew}. A crucial point about this
uncovering is that such a phantom
energy could well be the class of exotic matter which is required
to support traversable wormholes \cite{Sushkov:2005kj,Lobo:2005us}. This result could be regarded
to be one of the most powerful arguments in favor of the idea that
wormholes should no longer be regarded as just describing purely
mathematical toy space-time models with interest only for science
fictions writers, but also as plausible physical realities that
could exist in the very spacetemporal fabric of our Universe. The
realization of this fact has motivated a renaissance of the study
of wormhole space-times, the special interest being the
consideration of the possible accretion of phantom energy onto
wormholes, which may actually cause the growth of the their mouths
\cite{GonzalezDiaz:2004vv}.

It is one of the main aims of the present paper considering the
potential thermodynamical properties of Lorentzian traversable in principle wormholes. That
idea can be motivated by the fact that even though the definition
of an event horizon is no longer possible for such wormholes,
actually one can still introduce for them the concept of a
trapping horizon. Now, Hayward has emphasized the importance
of characterizing black holes themselves through local
considerations \cite{Hayward,Hayward:1994bu,Hayward:1997jp,Hayward:2004fz}, given that
a global property like the event
horizon can not be tested by observers. In this spirit, he
constructed a formalism able to describe the thermodynamical
properties of spherical, dynamical black holes based on the
existence of trapping horizons. It follows then that the presence of trapping horizons in wormhole spacetimes  would also allow such wormholes to be studied and shows the similar
nature of black- and worm-hole \cite{Hayward:1998pp}, when both are characterized locally
by means of outer trapping horizons that are spacelike or null and
timelike, respectively. Therefore, the idea that wormholes
may show some characteristics and properties which paralleled those already found in
black holes, seems to be quite natural, including in particular a
wormhole thermodynamics (as already suggested and later introduced
by Hayward \cite{Hayward:1998pp}) as well as a possible Hawking-like thermal
radiation for wormholes which we will discuss in this paper.

Actually, a key idea of the present work is to note that the results
relating to the accretion phenomenon \cite{GonzalezDiaz:2004vv}, based in the Babichev et al. method \cite{Babichev:2004yx}, must complement those
which can be extracted from the Hayward formalism \cite{Hayward,Hayward:1994bu,Hayward:1997jp,Hayward:2004fz}, so allowing a
deeper understanding of the wormhole thermodynamics. That
understanding, together with some results about phantom
thermodynamics \cite{GonzalezDiaz:2004eu}, would provide any possible Hawking-like radiation from
wormholes with a well defined physical meaning.

The present work is outlined as follows: In Sec. II we firstly summarize some basic ideas about the Morris-Thorne wormhole and, secondly, the key concepts of the Hayward formalism. Such a formalism is applied to the Morris-Thorne wormhole in Sec. III. In Sec. IV we introduce a consistent characterization for dynamical wormholes, which will allow us to derive a thermal radiation and formulate a whole thermodynamics in Sec. V. Finally, in Sec.VI, the conclusions are summarized and further comments are added.

\section{Preliminaries.}
\subsection{The Morris-Thorne wormholes.}

Morris and Thorne \cite{Morris:1988cz} worked out a general,
static and spherically symmetric wormhole, which is both traversable and stable. Such a solution
describes a throat connecting two asymptotically flat regions of
the spacetime, without the presence of any event horizon, that is
\begin{equation}\label{uno}
 {\rm d}s^2=e^{2\Phi(l)}{\rm d}t^2+{\rm d}l^2+r^2(l)\left[{\rm d}\theta^2+
 \sin^2{\rm d}\varphi^2\right],
\end{equation}
where the coordinate $-\infty<l<\infty$ and $\Phi(l)$ must be
finite everywhere. The radius of the wormhole throat is the
minimum of the function $r(l)$, $r_0$, which we can suppose,
without loss of generality, placed at $l=0$; therefore $l<0$ and
$l>0$ respectively cover the two asymptotically flat regions connected through
the throat at $l=0$. Note that the limit of $r(l)/|l|$ and
$\Phi(l)$ when $l\rightarrow\pm\infty$ must equal unity and
constant, respectively, in order to have an asymptotically flat
spacetime. Although this spacetime can be easily interpreted by
constructing an embedding diagram, it is also useful to express
metric (\ref{uno}) in terms of Schwarzschild coordinates. So we have
\begin{equation}\label{dos}
 {\rm d}s^2=e^{2\Phi(r)}{\rm d}t^2+\frac{{\rm d}r^2}{1-K(r)/r}+
 r^2\left[{\rm d}\theta^2+\sin^2{\rm d}\varphi^2\right],
\end{equation}
where two coordinate patches are needed now to cover the two
asymptotically flat regions, each with $r_0\leq r\leq\infty$. In
Eq. (2) $\Phi(r)$ and $K(r)$ are the redshift function and the
shape function, respectively in order to preserve asymptotic flatness, both such functions must tend to a constant
value when the radial coordinate goes to infinity. The minimum radius $r_0$ corresponds to the throat, where $K(r_0)=r_0$ and the embedded surface is vertical. The proper radial distance is
expressed by
\begin{equation}\label{tres}
 l(r)=\pm\int^r_{r_0}\frac{{\rm d}r^*}{\sqrt{1-K(r^*)/r^*}},
\end{equation}
and must be finite throughout the whole spacetime, i. e.,
$K(r)/r\leq1$.

Some constraints of interest regarding the exotic nature of the
background material which supports this spacetime can be deduced.
The Einstein equations imply that the sign of the radial
derivative of the shape-function, $K'(r)$, should be the same as
the sign of the energy density $\rho(r)$; in order to minimize the exoticity, therefore, it could be
advisable to demand $K'(r)>0$.
On the other hand, the embedding surface must flare outward, so
that $K'(r)<K(r)/r$ at or near the throat. This implies
$p_r(r)+\rho(r)<0$ on this regime, where $p_r(r)$ is the radial
pressure of the material. So, the exotic matter which supports this
wormhole spacetime violates the null energy condition, and this is of course a rather uncomfortable characteristic, at least classically. Nevertheless,
some quantum effects, such as the Casimir effect, have been shown to
violate this condition in nature, too.

As we have already mentioned in the Introduction, the above
properties of exotic matter have undergone a certain remake and gaining of
naturalness, originating from the discovery of the accelerated
expansion of the Universe. In fact, such an accelerated expansion
may very plausibly be explained by the existence of a fluid
homogeneously distributed throughout the universe, known as
phantom energy \cite{Caldwell:1999ew}, with an equation of state $p=w\rho$ in which $w>-1$.
Suskov \cite{Sushkov:2005kj} and Lobo \cite{Lobo:2005us} have extended the notion of phantom
energy to inhomogeneous spherically symmetric spacetimes by
regarding that the pressure related to the energy density through
the equation of state parameter must be the radial pressure,
calculating the transverse components by means of the Einstein
equations. One can see \cite{Lobo:2005us} that a particular specification
of the redshift and shape functions in metric (\ref{dos}) leads to a
static phantom traversable wormhole solution (where no-dynamical evolution for
the phantom energy is considered) which satisfies the
above-mentioned conditions, in particular the outward flaring
condition $K'(r_0)<1$.

\subsection{Trapping horizons.}

In this subsection we summarize some of the concepts and the notation
used in the Hayward formalism based on the dual-null dynamics
applicable to spherical symmetric solutions \cite{Hayward:1997jp}. The
line-element of a spacetime with spherical symmetry can be
generally written as
\begin{equation}\label{cuatro}
{\rm d}s^2=2g_{+-}{\rm d}\xi^+{\rm d}\xi^-+r^2{\rm d}\Omega^2,
\end{equation}
where $r$ and $g_{+-}$ are functions of the null coordinates
$(\xi^+,\xi^-)$, related with the two preferred null normal directions of each symmetric sphere $\partial_{\pm}\equiv\partial/\partial\xi^{\pm}$, $r$ is the so-called areal radius \cite{Hayward:1997jp} and ${\rm d}\Omega^2$
refers to the metric on the unit two-sphere. One can define the
expansions as
\begin{equation}\label{cinco}
\Theta_{\pm}=\frac{2}{r}\partial_{\pm}r,
\end{equation}
Since the sign of
$\Theta_{+}\Theta_-$ is invariant, one can say that a sphere is trapped,
untrapped or marginal if the product $\Theta_+\Theta_-$ is bigger,
less or equal to zero, respectively. If the orientation $\Theta_+>0$ and $\Theta_-<0$ is locally fixed on an untrapped sphere, then $\partial_+$ and $\partial_-$ are also fixed as the outgoing and ingoing null normal vectors (or the contrary if the orientation $\Theta_+<0$ and $\Theta_->0$ is considered). A marginal sphere with
$\Theta_+=0$ is future if $\Theta_-<0$, past if $\Theta_->0$ and
bifurcating\footnote{It must be noted that on the first part of this work we will consider future and past trapping horizons with $\Theta_+=0$, implying that $\xi^-$ must be related to the ingoing or outgoing null normal direction for future ($\Theta_-<0$) or past ($\Theta_->0$) trapping horizons, respectively.} if $\Theta_-=0$. This marginal sphere is outer if $\partial_-\Theta_+<0$,
inner if $\partial_-\Theta_+>0$ and degenerate if
$\partial_-\Theta_+=0$. A hypersurface foliated by marginal
spheres is called a trapping horizon and has the same
classification as the marginal spheres.

In spherical symmetric spacetimes a unified first law of thermodynamics can be
formulated \cite{Hayward:1997jp}. The gravitational energy in spaces with this
symmetry is the Misner-Sharp energy \cite{Misner:1964je} which can be defined by
\begin{equation}\label{energia1}
 E=\frac{1}{2}r\left(1-\partial^ar\partial_ar\right)=\frac{r}{2}\left(1-2g^{+-}\partial_+r\partial_-r\right).
\end{equation}
This expression  obviously must become $E=r/2$ on a trapping horizon (for properties {\it of $E$} see \cite{Hayward:1993ph}).

On the other hand, there are two invariants constructed out of the energy-momentum tensor of the background fluid which can be easily
expressed in these coordinates:
\begin{equation}\label{seis}
 \omega=-g_{+-}T^{+-}
\end{equation}
and the vector
\begin{equation}\label{siete}
\psi=T^{++}\partial_+r\partial_+ +T^{--}\partial_-r\partial_-.
\end{equation}
Then the first law of thermodynamics can be written in the following form
\begin{equation}\label{ocho}
\partial_{\pm}E=A\psi_{\pm}+\omega\partial_{\pm}V,
\end{equation}
where $A=4\pi r^2$ is the area of the spheres of symmetry, which is a geometrical-invariant, and $V=4\pi r^3/3$ is defined as the corresponding flat-space volume, dubbed as areal volume by Hayward \cite{Hayward:1997jp}. The first
term in the r.h.s. could be interpreted as an energy-supply term, i. e., this
term produces a change in the energy of the spacetime due to the
energy flux $\psi$ generated by the surrounding material (which
generates this geometry). The second term, $\omega\partial_{\pm}V$, behaves like a work
term, something like the work that the matter content must do to
support this configuration.

The Einstein equations of interest in terms the above coordinates
are
\begin{equation}\label{E1}
\partial_{\pm}\Theta_{\pm}=-\frac{1}{2} \Theta^2_{\pm}-\Theta_{\pm}\partial_{\pm}{\rm log}\left(-g_{+-}\right)
-8\pi T_{\pm\pm},
\end{equation}
\begin{equation}\label{E2}
\partial_{\pm}\Theta_{\mp}=-\Theta_+\Theta_-+\frac{1}{r^2}g_{+-}+8\pi T_{+-}.
\end{equation}

On the other hand, Kodama \cite{Kodama:1979vn} introduced a vector $k$ which can be understood as a generalization from the stationary Killing vector in spherically symmetric spacetimes, reducing to it in the vacuum case.\footnote{It must be emphasized that we have referred to the vacuum case, not to the more general static case.} That vector can be defined as
\begin{equation}
k={\rm curl}_2r,
\end{equation}
where the subscript 2 means referring to the two-dimensional
space normal to the spheres of symmetry.
One can express $k$ in terms of the null-coordinates yielding
\begin{equation}\label{nueve}
k=-g^{+-}\left(\partial_+r\partial_- -\partial_-r\partial_+\right),
\end{equation}
with norm
\begin{equation}
||k||^2=\frac{2E}{r}-1.
\end{equation}
This vector provides the trapping horizon with the additional definition of a
hypersurface where the Kodama vector is null. Therefore, such as
it happens in the case of static spacetimes, where a boundary can be
generally defined as the hypersurface where the temporal Killing vector is null, in the present case we must instead use the Kodama vector.
It can be seen that this vector generates a preferred
flow of time and that $E$ is its Noether charge. Moreover, $k$ has some
special properties \cite{Hayward:1994bu} similar to those of a Killing
vector on a static spacetime with boundaries, so allowing the
definition of a generalized surface gravity, given by
\begin{equation}\label{diez}
\kappa=\frac{E}{r^2}-4\pi r \omega,
\end{equation}
satisfying
\begin{equation}
k\cdot\left(\nabla\wedge k^b\right)=\pm\kappa k^b\, \, {\rm on\, a\, trapping\, horizon.}
\end{equation}
It is worth noticing at this stage that the generalized surface gravity can be equivalently expressed as
\begin{equation}\label{sg2}
\kappa=\frac{1}{2}g^{ab}\partial_a\partial_br.
\end{equation}
It follows that outer, degenerate and inner trapping horizons have $\kappa>0$, $\kappa=0$ and $\kappa<0$, respectively.

The Kodama vector was initially introduced in this formalism by
Hayward as an analogue of the temporal Killing vector in dynamical, spherically symmetric spacetimes. Such a vector gets even further interest in the case of wormhole spacetimes. Although there is
a temporal Killing vector for static traversable wormholes, it is not
vanishing everywhere, so precluding the introduction of a Killing
horizon, where one could try to define a surface gravity. However,
the definition of the Kodama vector and whereby the existence of a
trapping horizon imply the existence of the
generalized surface gravity in static and dynamic traversable wormholes, as we
will see in the next section.

Finally, one can project Eq.(\ref{ocho}) along the trapping horizon to obtain
\begin{equation}\label{ochob}
L_zE=\frac{\kappa L_zA}{8\pi}+\omega L_zV,
\end{equation}
where $L_z=z\cdot\nabla$ and $z=z^+\partial_+ +z^-\partial_-$ is tangent to the trapping horizon. This expression allows us to introduce a relation between the geometric entropy and the surface area
\begin{equation}\label{entropia}
S\propto A|_H.
\end{equation}

\section{Hayward Formalism applied to Morris-Thorne wormholes.}
The Hayward formalism was introduced for defining the
properties of real black holes in terms of measurable quantities.
It allows for a formulation of the thermodynamics of dynamical
black holes which consistently recovers the results obtained by
global considerations using the event horizon in the vacuum static case.
It is in this sense that such a formalism can be regarded as a
generalization, provided that the local quantities are physically
meaningful both in static and dynamical spacetimes. A nice and
rather surprising feature appears when one realizes that the
Morris-Thorne wormhole (where it is not possible to infer any
property similar to those found in black hole physics using global
considerations) shows thermodynamical characteristics similar to
those of black holes if local quantities are considered. That can
be better understood if one notices that the Schwarzschild black hole is
the only spherically symmetric solution in the vacuum, and
therefore, any dynamical generalizations of black holes must be
formulated in the presence of some matter content. The maximal extension
of the Schwarzschild spacetime can be understood as a pair black-white hole, or an
Einstein-Rosen bridge, which corresponds to a vacuum solution
which describes a wormhole and has
associated a given thermodynamics. Nevertheless, the Einstein-Rosen bridge can not be traversed since it has an event horizon. If we consider wormholes which
in principle are traversable even in the static case, that is
Morris-Thorne wormholes, some matter content must be present to
support their structure. So the necessity of a formulation in terms
of local quantities, measurable for an observer with finite life,
must be related to the presence of some matter content, rather
than with a dynamical evolution of the spacetime. In this section
we apply the results obtained by Hayward for spherically symmetric
solutions to static wormholes and rigorously show their
consequences, some of which were already suggested and/or
indicated by Ida and Hayward himself \cite{Ida:1999an}.

In order to re-express metric (\ref{dos}) in the form given by Eq.
(\ref{cuatro}), a possible choice of the null-coordinates results from
taking $\xi^+=t+r_*$ and $\xi^-=t-r_*$, with $r_*$ such that ${\rm
d}r/{\rm d}r_*=\sqrt{-g_{00}/g_{rr}}=e^\Phi(r)\sqrt{1-K(r)/r}$,
and $\xi^+$ and $\xi^-$ being respectively related with the
outgoing and ingoing radiation. By the definitions introduced in
the previous section, one can readily note that there is a
bifurcating trapping horizon at $r=r_0$, which is outer whenever
$K'(r_0)<1$ (which must be satisfied by the outward flaring
condition).

The Misner-Sharp energy, ``energy density'' and ``energy flux''
are respectively given by
\begin{equation}
E=\frac{K(r)}{2},
\end{equation}
\begin{equation}
\omega=\frac{\rho-p_r}{2}
\end{equation}
and
\begin{equation}
\psi=-\frac{\left(\rho+p_r\right)}{2}e^{-\Phi(r)}\sqrt{1-K(r)/r}(-1,1).
\end{equation}
It can be noticed that $E=r/2$ only at the throat (trapping
horizon) and that, as in the case of Sushkov \cite{Sushkov:2005kj} and Lobo \cite{Lobo:2005us}, no information about the transverse components of the
pressure is needed. Now, Eq. (\ref{ocho}) for the first law can be particularized to the Morris-Thorne case whenever we introduce the following quantities
\begin{equation}\label{varE}
\partial_{\pm}E=\pm 2\pi r^2\rho e^\Phi\sqrt{1-K(r)/r}
\end{equation}
and
\begin{equation}\label{varpsi}
\psi_{\pm}=\pm e^\Phi(r)\sqrt{1-K(r)/r}\frac{\rho+p_r}{4}.
\end{equation}
Obviously, all terms in
Eq.(\ref{ocho}) should vanish at the horizon in the static case, i.e., the throat of
the Morris-Thorne wormhole has no dynamical evolution. When
compared with those of the black hole case, however, the study of
the above quantities could give us some deeper understanding of such a
spacetime which is based on the presence of the exotic matter. We
stress first of all that the variation of the gravitational energy
would always be positive in the outgoing direction and negative in
the ingoing direction because\footnote{It is worth noting
that $e^\Phi\sqrt{1-K(r)/r}\equiv\alpha$ appears by explicitly considering the
Morris-Thorne solution,
where $\alpha=\sqrt{-g_{00}/g_{rr}}$ is a general factor at least in spherically symmetric, static cases; therefore it
has the same sign as in the case of a static black hole surrounded by ordinary matter without dynamical evolution. We will assume that in the dynamical cases of black and worm-holes $\alpha$ keeps the sign unchanged relative to the one appearing in Eqs.(\ref{varE}) and (\ref{varpsi}) outside the throat.} $\rho>0$. It follows that
in this case the exoticity or not exoticity of the matter content
is not important for the result. The ``energy density'' $\omega$,
which is non-negative for static or dynamic black holes, actually
is positive even though the energy conditions are not fulfilled.
The key differentiating point appears when one considers the
energy-supply term because the energy flux depends on the sign of
$\rho+p_r$; whereas in case of usual ordinary matter that could
be interpreted as a fluid which ``gives'' energy to the spacetime,
for exotic matter the fluid ``receives'' or ``gets'' energy from
the spacetime. It must be noted that the ``energy removal'', induced by the energy flux term in the wormhole case, can never be large enough to make
the variation of the gravitational energy changes sign.

The spacetime (\ref{dos}) possesses a temporal Killing vector
which is non-vanishing everywhere, and therefore it makes no sense
trying to find a surface gravity as used by Gibbons and Hawking \cite{Gibbons:1977mu} for
Killing horizons. However, spherical symmetry would still allow us
to consider the Kodama vector with components
\begin{equation}
k^{\pm}=e^{-\Phi(r)}\sqrt{1-K(r)/r},
\end{equation}
with $||k||^2=0$ at the throat. It follows that a generalized surface
gravity as defined in Eq.(\ref{diez}) can yet be obtained
\begin{equation}
\kappa|_H=\frac{1-K'(r_0)}{4r_0},
\end{equation}
where ``$|_H$'' means evaluation at the horizon. It can be noted
that, since the throat is an outer trapping horizon ($K'(r_0)<1$),
the surface gravity is positive, as it should be expected from the very definition of surface gravity (\ref{sg2}).

It is well known that when the surface gravity is defined by the use of a temporal Killing vector, this quantity is understood to mean that there is a force acting on test particles in a gravitational field. The generalized surface gravity is in turn defined by the use of the Kodama vector which determines a preferred flow of time, reducing to the Killing vector in the vacuum case and recovering the surface gravity its usual meaning. However, in the case of a spherically symmetric and static wormhole one can define both, the temporal Killing and the Kodama vector, being the Kodama vector of greater interest since it vanishes at a particular surface. Moreover, in dynamical spherically symmetric cases one can only define the Kodama vector. Therefore we could suspect that the generalized surface gravity should originate some effect on test particles which would go beyond that corresponding to a force, and only reducing to it in the vacuum case. On the other hand, if by some kind of symmetry this effect on a test particle would vanish, then we should think that such a symmetry would also produce that the trapping horizon be degenerated.

\section{Dynamical wormholes.}
The existence of a well-defined surface gravity and the possible
identification of the first term in the r.h.s of Eq. (\ref{ochob})
to something proportional to an entropy seems to suggest the possibility that a
proper wormhole thermodynamics can be formulated, as it was
already commented in Ref. \cite{Hayward:1998pp}. Nevertheless, a
more precise definition of the dynamical wormhole horizon must be done in
order to settle down univocally its characteristics. To get that aim we
first have to comment on the results obtained for the increase of the
black hole area \cite{Hayward:2004fz}, comparing then them with those derived from the
accretion method \cite{Babichev:2004yx}, which would be expected to lead to additional information in
the case of wormholes.

First, the area of a surface can be expressed in terms of the area
2-form $\mu$ as $A=\int_S\mu$, with $\mu=r^2\sin\theta{\rm
d}\theta{\rm d}\varphi$ in the spherically symmetric case.
Therefore, one can study the evolution of the area by using
\begin{equation}\label{area}
L_zA=\int_s\mu\left(z^+\Theta_++z^-\Theta_-\right),
\end{equation}
with $z$ tangent to the considered surface.

On the other hand, by the very definition of a trapping horizon we
can fix $\Theta_+|_H=0$, which leads to a description of the
evolution of the horizon through
\begin{equation}\label{expansion}
L_z\Theta_+|_H=\left[z^+\partial_+\Theta_++z^-\partial_-\Theta_+\right]|_H=0.
\end{equation}

Then, one can also note that by evaluating Eq.(\ref{E1}) at the
trapping horizon we can obtain
\begin{equation}\label{tipomateria}
\partial_+\Theta_+|_H=-8\pi T_{++}|_H,
\end{equation}
with $T_{++}\propto \rho+p_r$. Therefore, if it is a ordinary matter
that is supporting the spacetime, then one has
$\partial_+\Theta_+|_H<0$ but if we have exotic matter instead,
then $\partial_+\Theta_+|_H>0$.

Characterizing black holes by the presence of a future outer trapping horizon
implies the growth of its area, provided that its
matter content satisfies the classical dominant energy conditions \cite{Hayward:2004fz}. This is
easily seen by taking into account the definition of outer trapping
horizon and by realizing that, when introduced in condition
(\ref{expansion}), Eq.(\ref{tipomateria}), for ordinary matter,
implies that the sign of $z^+$ and $z^-$ must be different, i.e.
outer trapping horizons are achronal in the presence of ordinary
matter. Then, evaluating $L_zA$ at the horizon, taking into
account that the black hole horizon is future and considering that $z$ has a
positive component along the future-pointing direction of
vanishing expansion, $z^+>0$, one finally obtains that\footnote{In the case of white holes, we have a past
outer trapping horizon and therefore $L_zA\leq0$.} $L_zA\geq0$. It is worth
noticing that when exotic matter is accreted, then the previous
reasoning would imply that the black hole area must decrease.

Whereas the characterization of black holes following the above lines
appears to be rather natural, a reasonable doubt may still be kept about how
the outer trapping horizon of wormholes may be considered. Since a traversable wormhole should necessarily be described in the presence of exotic matter, the above
considerations imply that its trapping horizon should be
timelike. However, if the timelike trapping horizon is future (past) then, by Eq. (\ref{area}), its area
would decrease (increase). Of course its area should remain constant in the
static case, as in that case the horizon must be bifurcating.

A consistent choice between these two possibilities can be done by recalling what happens in the
process of dark energy accretion onto black holes, Ref. \cite{Babichev:2004yx}. In fact, that
process leads to an unavoidable decrease (increase) of their area if $p+\rho<0$ ($p+\rho>0$), where $p$ could be identified to $p_r$ in our case. Since the phantom
energy can be considered as a particular case of exotic matter
\cite{Sushkov:2005kj,Lobo:2005us}, we recover again the same result as in the precedent paragraph, that is, both formalism can be shown to describe the same mechanism. The
method of Babichev et al., Ref. \cite{Babichev:2004yx}, can also be applied to wormholes
\cite{GonzalezDiaz:2004vv}, obtaining that the size of the
wormhole throat must increase by the accretion of phantom energy
and decrease if $p+\rho>0$. That result could have been
suspected from the onset since, if the phantom energy can be regarded as the kind
of matter which supports wormholes, it becomes quite a reasonable
implication that the wormhole throat must decrease in size if
matter with the opposite internal energy is accreted, getting, on
the contrary, on bigger sizes when phantom energy is accreted. It follows that
one must characterize dynamical wormholes by their past outer trapping
horizon, so implying that $L_zA>0$.

It also follows then that, whereas the characterization of a black hole
must be done in terms of its future trapping horizon (as such a black
hole would tend to be static as one goes into the future, ending
its evolution at infinity), that of a white hole, which is assumed
to have born static and then allowed to evolve, should be done in
terms of its past horizon. Thus, in the case of a wormhole one can
consider a picture of it being born at the beginning of the
universe (or when an advanced civilization constructs it at some
later moment) and then left to evolve, increasing or
decreasing its size when there is a higher proportion of exotic or ordinary matter,
respectively. Therefore, taking the past horizon to characterize the wormholes
appears to be consistent with this picture.

The above mentioned conclusion is in
agreement as well with the idea that phantom energy is equivalent
to usual energy traveling backward in time, which seems to somehow
imply that a wormhole spacetime supported by phantom energy could
be physically equivalent to a black hole spacetime supported by
ordinary matter defined in an reversed time. In fact, a time
reversed future horizon would necessarily become a past horizon.
It must be emphasized that in this argument we are considering black holes surrounded by ordinary matter. Static black holes can be purely vacuum solutions which are invariant under time reversal. But these vacuum black holes would be compared to purely vacuum wormholes, which, in turn, should also be invariant under time reversal. In particular, and since we are concerned with spherically symmetric spacetimes, one could consider the Einstein-Rosen bridge, which is a purely vacuum (non-traversable) wormhole; that wormhole is not only invariant under time reversal, but it has the same properties as Schwarzschild black hole since it is nothing but an extension of that spacetime. So, we can conclude that the time invariance should be broken when any kind of matter is present, i. e., when we have a non-vanishing r.h.s in Einstein equations.

On the other hand, taking into account the proportionality relation (\ref{entropia}) we can see that the dynamical evolution of the wormhole entropy must be such that $L_zS\geq0$, which saturates only at the static case.

\section{Wormhole thermal radiation and thermodynamics.}

The existence of a non-vanishing surface gravity at the wormhole
throat seems to imply that it can be characterized by a non-zero
temperature so that one would expect that wormholes should emit some sort of thermal
radiation. Of course, we are considering two-way traversable
wormholes and, therefore, any matter or radiation could pass
through it from one universe to an other (or from a region
to another of a single universe) finally appearing in the latter
universe (or region). What we are refereeing to, nevertheless, is a
completely different kind of radiative phenomenon. One can distinguish that
phenomenon from that implying a
radiation traveling through the wormhole because whereas the
latter phenomenon follows a path which is allowed by classical
general relativity, thermal radiation is an essentially
quantum-mechanical process. Therefore, even in the case that no
matter or radiation would travel through the wormhole classically,
the existence of a trapping horizon would produce a quantum
thermal radiation.

In Ref. \cite{Hayward:2008jq} the use of a Hamilton-Jacobi
variant of the Parikh-Wilczek tunneling method led to a local
Hawking temperature in the case of spherically symmetric black
holes. It was however also suggested \cite{Hayward:2008jq} that this method was only
applicable to a future outer trapping horizon, since non physical
meaning could be addressed to the case of horizons
other than the past inner trapping horizons. Notwithstanding, although the past outer trapping horizon belongs to the first of the latter cases, it actually describes physically allowable dynamical wormholes. Thus, in this
section we apply the method considered in Ref. \cite{Hayward:2008jq} to
past outer trapping horizon and show that the result advanced in
the precedent paragraph is physically consistent, provided we take
into account the pre-suppositions and results derived \cite{GonzalezDiaz:2004eu} for
phantom energy thermodynamics.

Let us consider a dynamical, spherically symmetric wormhole,
described by the metric given in Eq. (\ref{cuatro}), having a past
trapping horizon with $\Theta_-=0$ and\footnote{One could equivalently (following the notations and calculations used so far in this work) take $\Theta_+=0$ and $\Theta_->0$, and then $\xi^+$ ($\xi^-$) would describe ingoing (outgoing) radiation.} $\Theta_+>0$. In order to
study any possible thermal radiation emitted by that spacetime
tunnel, we can express metric (\ref{cuatro}) in terms of the generalized
retarded Eddington-Finkelstein coordinates
\begin{equation}\label{EF}
{\rm d}s^2=-e^{2\Psi}C{\rm d}u^2-2e^\Psi{\rm d}u{\rm d}r+r^2{\rm
d}\Omega^2,
\end{equation}
where $u=\xi^-$ and ${\rm d}\xi^+=\partial_u\xi^+{\rm
d}u+\partial_r\xi^+{\rm d}r$. Since $\partial_r\xi^+>0$, we have
considered $e^\Psi=-g_{+-}\partial_r\xi^+$ and
$e^{2\Psi}C=-2g_{+-}\partial_u\xi^+$, with $C=1-2E/r$, $E$ being defined
by Eq. (\ref{energia1}), and $\Psi$ expressing the gauge freedom in
the choice of the null coordinate $u$. The use of retarded
coordinates ensures that the marginal surfaces, for which $C=0$,
actually are past marginal surfaces.

Now, similarly to as it has been done in Ref.\cite{Hayward:2008jq}, we consider a massless scalar
field in the eikonal approximation,
$\phi=\phi_0\exp\left(iI\right)$, with a slowly varying amplitude
and a rapidly varying action given by
\begin{equation}\label{accion}
I=\int \omega_\phi e^\Psi{\rm d}u-\int k_\phi{\rm d}r,
\end{equation}
which, in our case, actually describes radially outgoing
radiation. Infalling radiation would in fact require the use of
advanced coordinates instead. The wave equation $\nabla^2\phi=0$
in turn implies
\begin{equation}
g^{ab}\nabla_aI\nabla_bI=0,
\end{equation}
which, taking into account that $\partial_uI=e^\Psi\omega_\phi$
and $\partial_rI=-k$, leads to
\begin{equation}\label{k}
k_\phi^2C+2\omega_\phi k_\phi=0.
\end{equation}
One solution of Eq. (\ref{k}) is $k=0$. This must correspond to
outgoing modes, as we are considering that $\phi$ is
outgoing. Therefore, the alternate solution,
$k=-2\omega_\phi/C<0$, should correspond to the ingoing modes.
Since $C$ vanishes on the horizon, the action (\ref{accion}) has a
pole. Noting that $\kappa|_H=\partial_rC/2$, where, without loss of generality, we have taken $\Psi=0$. Given that $\kappa|_H$ should be gauge-invariant, with
expanding $C$, one obtains
$k_\phi\approx-\omega_\phi/\left[\kappa|_H(r-r_0)\right]$.
Therefore the action has an imaginary contribution which is
obtained deforming the contour of integration in the upper $r$
half-plane
\begin{equation}\label{imaccion}
{\rm Im}\left(I\right)|_H=-\frac{\pi\omega_\phi}{\kappa|_H}.
\end{equation}
One can take the particle production rate as given by the
WKB approximation of the tunneling probability $\Gamma$ along a
classically forbidden trajectory
\begin{equation}\label{prob}
\Gamma\propto\exp\left[-2{\rm Im}\left(I\right)\right].
\end{equation}
Although the traversing path of a traversable wormhole is a
classically allowed trajectory and that wormhole is interpretable as
a two-way traversable membrane, we can consider that the existence
of a trapping horizon opens the possibility for an additional quantum traversing phenomenon through the wormhole. This additional quantum radiation would be somehow based on some sort of quantum tunneling mechanism between the two
involved universes (or the two regions of the same, single
universe), a process which of course is classically
forbidden. If such an interpretation is accepted, then (\ref{prob}) takes
into account the probability of particle production rate at the
trapping horizon induced by some quantum, or at least semi-classical, effect. Considering probability (\ref{prob}), the
resulting thermal form $\Gamma\propto\exp\left(-\omega_\phi/T_H\right)$, would lead us to finally compute
a temperature for the thermal radiation given by
\begin{equation}\label{temperature}
T=-\frac{\kappa|_H}{2\pi},
\end{equation}
which is negative. At first sight, one could think that we could
save ourselves from this negative temperature because it is related to the
ingoing modes. However this can no longer be the case as this thermal
radiation appears at the horizon, so again leading to a negative horizon
temperature.

Even if one tries to avoid any possible physical
implication for this negative temperature, claiming that it
would be only a problem if one is at the horizon, one must take
into account that the infalling radiation getting in one of the
wormhole mouths would travel through that wormhole following a classical path to re-appear at the other mouth as an outgoing radiation in the
other universe (or the other region of universe). Now, since the
same phenomenon would take place at both mouths, in the end of the
day it would be fully unavoidable that one finds outgoing
radiation with negative temperature in the universe whenever a wormhole
is present.

Moreover, as we have already discussed, phantom energy must
be considered as a special case of exotic matter, and it is also
known that phantom energy must always be characterized by a
negative temperature \cite{GonzalezDiaz:2004eu}. Therefore, the above results should be taken to be a consistency proof of the used method as a negative radiation temperature simply
express the feature to be expected that wormholes should emit a
thermal radiation just of the same kind as that of the stuff
supporting them, such as it also occurs with dynamical black holes
with respect to ordinary matter and positive temperature. On the
other hand, as we have indicated at the end of Sec. IV, phantom
energy can be treated as ordinary matter traveling backward in time,
and the wormhole horizon as a black hole horizon for reversed
time, at least if we are considering the real, dynamical, case.
One can also note that the retarded metric (\ref{EF}) which
allows us to derive the negative temperature, is nothing but the time
reversed line element obtained from the corresponding advanced
metric, that is the spacetime driving thermal radiation with
positive temperature in the case of dynamical black holes. In this
way, one could interpret this result by considering that only some
fundamental quantity, such as the entropy, must be invariant under
time reversal, but that quantities like temperature must change
sign under time reversal $t\rightarrow-t$. Even more, taking into
account (\ref{temperature}), one can
re-express Eq.(\ref{ochob}) in the form
\begin{equation}\label{leyuno}
L_zE=-TL_zS+\omega L_zV
\end{equation}
on the trapping horizon, so univocally defining the wormhole geometric entropy by
\begin{equation}
S=\frac{A|_H}{4}.
\end{equation}
A negative sign for the first term in the r.h.s. of Eq.(\ref{leyuno})
is not surprising since, as already pointed out in
Sec.III, the exotic matter which supports this spacetime gets
energy from the spacetime itself, so the term for the energy
supply must be negative. Following that line of reasoning we can
now formulate the first law of wormhole thermodynamics as:

{\it First law: The change in the gravitational energy of a
wormhole equals the sum of the energy removed from the
wormhole plus the work done in the wormhole.}

This can be interpreted by considering that the exotic matter is responsible for both the energy
removal and the work done, the balance always giving rise to a positive
variation of the total gravitational energy.

On the other hand, we have also shown in Sec. IV that $L_zA\geq0$ and
$L_zS\geq0$, which saturate only at the static case. If we assume
a real, cosmological wormhole created in a rather exotic dynamical background, we can formulate the second law for
wormhole thermodynamics as follows

{\it Second law: The entropy of a dynamical wormhole is given by its surface area which always
increases.}

Moreover, since by definition a wormhole can be characterized by a
past outer trapping horizon, no dynamical evolution whatsoever can
change the outer character of its horizon. In Ref.\cite{Hayward:1998pp} it was considered that a
wormhole could change its nature becoming a black hole, but
even in the case that this process would be possible, the trapping
horizon could never change from being outer or inner. This in turn
implies that no dynamical evolution can change the sign of the
surface gravity, so that we always have $\kappa>0$. Therefore, if
the the trapping horizon remains being past, Eq.
(\ref{temperature}) must be valid, and $T<0$. Now, from this
argument we can straightforwardly formulate the third law of wormhole thermodynamic
in two alternate ways

{\it Third law (first formulation): It is impossible to reach the
absolute zero for surface gravity by any dynamical process.}

{\it Third law (second formulation): In a wormhole it is
impossible to reach the absolute zero of temperature by any
dynamical process.}

It is worth noticing that if by using some line of reasoning similar to
that introduced in \cite{Hayward:1998pp}, it would be made possible to change the
horizon from being future to being past or vice versa, such a change would by equivalence between matter
and geometry imply
the change of the very nature of the equation of state of the surrounding
matter\footnote{It can be noted that the argument used in Ref.
\cite{Hayward:1998pp} about the possible conversion of a past outer trapping
horizon (i.e. a wormhole which is bounded by this horizon) into a future outer trapping horizon
(i.e. a black hole which is bounded by this horizon) and vice versa, are just based on an unavoidable previous
change of the surrounding matter equation of state,
bringing exotic matter into ordinary matter or vice versa. From the equivalence between the matter content and the geometry,
one can then rise the question: what is first, the egg or the
hen?}. Therefore, the first law of wormholes thermodynamics would
then become the first law of black holes thermodynamics, where the
energy is supplied by ordinary matter rather than by the exotic one
and the minus sign in Eq.(\ref{leyuno}) is replaced by a plus
sign. The latter implication arises from the feature that a future
outer trapping horizon should produce thermal radiation at a
positive temperature. The second law would remain then unchanged
since it can be noted that the variation of the horizon area, and
hence of the entropy, is equivalent for a past outer trapping
horizon surrounded by exotic matter and for a future outer
trapping horizon surrounded by ordinary matter. 

Finally, the two formulations provided for the third law would also be the same, but in the
second formulation one would consider that the temperature took
only on positive values.

\section{Conclusions and further comments.}

In this work we have applied results related to a generalized first
law of thermodynamics \cite{Hayward:1997jp} and the existence of a 
generalized surface gravity \cite{Hayward:1997jp,Ida:1999an} to the case of the Morris-Thorne wormholes \cite{Morris:1988cz}, where the outer trapping
horizon is bifurcating. Since these wormholes correspond to static
solutions, no dynamical evolution of the throat is of course
allowed, with all terms entering the first law vanishing at the
throat. However, the comparison of the involved quantities (such
as the variation of the gravitational energy and the energy-exchange so as work terms as well) with the case of black holes surrounded
by ordinary matter actually provide us with some useful information
about the nature of this spacetime (or alternatively about the exotic matter),
under the assumption that in the dynamical cases these quantities keep the signs unchanged relative to those appearing outside the throat in the static cases.
It follows that the variation of the gravitational energy and the
``work term'', which could be interpreted as the work carried out by
the matter content in order to maintain the spacetime, have the same sign in spherically
symmetric spacetimes supported by both ordinary and exotic matter. Notwithstanding, the
``energy-exchange term'' would be positive in the case of dynamical black
holes surrounded by ordinary matter (i. e. it is an energy supply) and negative for
dynamical wormholes surrounded by exotic
matter (i. e. it corresponds to an energy removal).

It must be also emphasized that the Kodama vector, which allows us to introduce a
generalized surface gravity in dynamic spherically symmetric
spacetimes \cite{Hayward:1997jp}, must be taken into account not only in the case of dynamical
solutions, but also in the more general case of non-vacuum
solutions. In fact, whereas the Kodama vector reduces to the temporal Killing in the spherically symmetric vacuum solution (Schwarzschild), that reduction is no longer possible for static non-vacuum solutions. That differentiation is a key ingredient in the case of wormholes, where even for the static Morris-Thorne case there is no
Killing horizon in spite of having a temporal Killing vector and possessing a non degenerate trapping horizon.
This in turn implies
a non-vanishing generalized surface gravity based on local
concepts which have therefore potentially observable consequences.

On the other hand, noting that the variation of
the black hole size obtained by using
the Hayward formalism \cite{Hayward:2004fz} for dynamical black holes is equivalent
to that variation coming form the consideration of dark energy accretion
onto black holes \cite{Babichev:2004yx}, we have assumed
that these two variations actually are alternate approaches in order to describe the same, unique phenomenon. When applied to wormholes such an identification leads to the characterization of such wormholes in terms of the
past outer trapping horizons.
That characterization would imply that the area (and hence the entropy) of a dynamical
wormhole always increases if there are no changes in the exoticity of
the background (second law of wormhole thermodynamics) and
that the hole appears to thermally radiate with negative temperature.
There must not be any problem
with the physical meaning of this radiation, because phantom energy
itself is also associated with a negative temperature. It is quite a natural conclusion that the wormhole would emit radiation of the same kind as the matter which supports it, such as it occurs in the case
of dynamical black hole evaporation with respect to ordinary matter.

These considerations allow us to consistently re-interpret the generalized first
law of thermodynamics as formulated by Hayward \cite{Hayward:1997jp} in the case of wormholes, noting that
in this case the change in the gravitational energy of the
wormhole throat is equal to the sum of the energy removed from the
wormhole and the work done on the wormhole (first law of wormholes thermodynamics), a result which is consistent with the above mentioned results obtained by analyzing of the Morris-Thorne spacetime.

It is worth noticing that because once the past outer trapping horizon is settled to characterize a wormhole, it will always remain outer, it becomes impossible that the surface gravity reaches an absolute zero along any dynamical processes (third law of wormhole thermodynamics). This result would imply that if the trapping horizon remains past, the wormhole throat would always have a temperature $T<0$, whereas if the trapping horizon would change to be future, then we had a black hole with positive temperature \cite{Hayward:2008jq}, the temperature never reaching a zero value in any case.

On the other hand, the idea that phantom energy can be
viewed as just ordinary matter subject to time inversion becomes fully
consistent whenever one takes into account that the dynamical wormhole
horizon could likewise be treated as the time reversed form of the
dynamical black hole horizon. In fact, the three laws of thermodynamical wormholes
reduces to the three laws of black hole dynamics when one
simply considers that time inversion renders exotic matter into ordinary
matter supplying energy to the spacetime, so changing sign
of temperature. It must be noted that in the purely vacuum case both, black hole and wormhole, should be invariant under time inversion with the r.h.s. of Einstein equations vanishing in both cases. In fact, such cases respectively correspond to e.g. the Schwarzschild black hole and to the Einstein-Rosen bridge. Nevertheless, the Einstein-Rosen bridge is a non-traversable wormhole; therefore, in order to compare a static traversable wormhole (Morris and Thorne) with a black hole, the latter solution must be static in the presence of ordinary matter. In such cases, although rigorously there is no dynamical evolution of the horizons, the correspondence between black hole and wormhole under time inversion can be already suspected from the onset.

We would like to remark that it is quite plausible that the existence of
wormholes be partly based on the possible presence of phantom
energy in our Universe. Of course, even though in that case the main
part of the energy density of the universe would be contributed by
phantom energy, a remaining $25\%$ would still be made up of ordinary matter
(dark or not). At least in principle, existing wormhole structures would be
compatible with the configuration of such a universe, even though
a necessarily sub-dominant proportion of ordinary matter be present,
provided that the effective equation of state parameter of
the universe be less than minus one.

At first sight, the above results might perhaps point out to a way
through which wormholes might be localized in our environment by simply
measuring the inhomogeneities implied phantom radiation,
similarly to as initially thought for black hole Hawking radiation
\cite{Gibbons:1977mu}. However, in both cases, at present, we are
far from having hypothetical instruments sensitive and precise
enough to detect any of the inhomogeneities and anisotropies which
could be expected from the thermal emission from black holes and
wormholes of moderate sizes.

After completion of this paper we became aware of a paper by
Hayward \cite{Haywardwh} in which some part of the present work were
also discussed following partly similar though somewhat divergent arguments.

\acknowledgments

\noindent P.~M.~M.
thanks B. Alvarez and L. Pastor for encouragement and gratefully
acknowledges the financial support provided by the I3P framework
of CSIC and the European Social Fund. This work was supported by
MEC under Research Project No.FIS2008-06332/FIS.

\end{document}